\documentclass{article}
\usepackage{spconf,amsmath,graphicx}
\usepackage{amsfonts}
\usepackage{subfigure}
\usepackage{hyperref}
\usepackage{booktabs,caption}
\usepackage[flushleft]{threeparttable}
\usepackage{environ} 

\title{High-quality Speech Synthesis Using Super-resolution Mel-Spectrogram}
\twoauthors
  {Leyuan Sheng, Dong-Yan Huang}
	{Speech Technology Department, \\UBTECH Robotics Corporate, \\Shenzhen, China}
  {Evgeniy N. Pavlovskiy}
	{Novosibirsk State University,\\ Novosibirsk, Russia}

\begin{document}
%
\maketitle
\begin{abstract}
In speech synthesis and speech enhancement systems, mel-spectrograms need to be precise in acoustic representations. However, the generated spectrograms are over-smooth, that could not produce high quality synthesized speech. Inspired by image-to-image translation, we address this problem by using a learning-based post f\mbox{}ilter combining Pix2PixHD and ResUnet to reconstruct the mel-spectrograms together with super-resolution. From the resulting super-resolution spectrogram networks, we can generate enhanced spectrograms to produce high quality
synthesized speech. Our proposed model achieves improved mean opinion scores (MOS) of 3.71 and  4.01 over baseline results of 3.29 and 3.84, while using vocoder Grif\mbox{}f\mbox{}in-Lim and WaveNet, respectively.
\end{abstract}
\section{Introduction}
Text to speech (TTS) synthesis aims at producing an intelligible and natural speech for a given text input. A traditional TTS system includes two parts: the front-end (text-analyzer) and the back-end (speech synthesizer). They usually consist of many domain-specif\mbox{}ic modules. The front-end requires a strong background in linguistics for text analysis and feature extraction.
The back-end needs to have a certain understanding of the vocal mechanism and signal processing of speech, from the features extracted by the front-end to complete the speech synthesis.

Recently, TTS approaches based on deep learning have been proposed and extensively investigated. Deep learning can integrate both the front-end and the back-end into a single ``end-to-end'' learning model, example of which are Tacotron \cite{tacotron}, Char2Wav \cite{char2wav}, Tacotron2 \cite{tacotron2} and ClariNet \cite{clarinet}. These end-to-end  generative  text-to-speech  models typically perform two tasks: 1) directly mapping the text into time-aligned acoustic features, as in Tacotron and Tacotron 2; 2) to convert the the generated acoustic features into a waveform  such as Grif\mbox{}f\mbox{}in-Lim algorithm \cite{griffin-lim}, WaveNet \cite{wavenet} and WaveGlow \cite{waveglow} is using a vocoder. It has been found that the mel-spectrograms representation is reasonable and ef\mbox{}fective for the use of speech synthesis systems \cite{tacotron} \cite{either2}. Therefore, we use mel-spectrograms to bridge the two steps in a TTS system. Note that after obtaining the mel-spectrograms in Tacotron and Tacotron2, there exists a post-net module to improve the mel-spectrograms predictions, which will improve the quality of synthesized speech. The closer the gap between the mel-spectrograms predicted by the model to the  mel-spectrograms extracted from ground truth audio, the higher the quality of synthesized speech.

As the research in generative model progress, new advanced Generative Adversarial Networks (GANs)  have started to emerge in speech processing. A speech enhancement GAN (SEGAN) in \cite{segan} and a Convolutional Neural Network (CNN)-based GAN in \cite{cnn-gan} was proposed for SE.  A GAN-based post-filter in \cite{either2} and \cite{stft2}  was proposed  for short-term Fourier transform (STFT) spectrograms. Moreover, the conditional GAN (cGAN) architecture adopted a pixel-to-pixel (Pix2Pix) framework has investigated for SE \cite{sesv} and Music Information Retrieval (MIR) \cite{either1}. As the high-resolution image synthesis architecture Pix2PixHD \cite{pix2pixhd} improves the performance than Pix2Pix, in \cite{ours} the authors have implemented the Pix2PixHD within a cGAN framework for reducing over-smoothness in speech synthesis.

In this work, we proposed a novel model to improve mel-spectrograms prediction for high-quality speech synthesis  by combining the advantages of Pix2PixHD \cite{pix2pixhd} and deep residual U-Net (ResUnet) \cite{resunet}.
Meanwhile, although several researchers have investigated models translating images to speech. But these either adapt the image translation model to modify the acoustic feature inputs \cite{either2} \cite{either1} \cite{either3}, or modify the image translation model itself \cite{or}. In practice, the image translation model is combined with acoustic feature processing.
These contributions of this paper are: 1) we demonstrate that the spectrogram can be viewed completely as an image, so that the task of speech signal processing can be processed by image restoration or translation model. 2) our proposed model can effectively improve speech quality based on generated enhanced spectrogram images.
\section{Related Work}

Our proposed super-resolution mel-spectrogram is based on the pix2pixHD conditional GAN-based image translation framework. In contrast to the original pix2pixHD model, the generator is used in a local enhancer network using ResUnet instead of the generator in GAN. In this section, we f\mbox{}irst review the conditional GANs followed and briefly describe pix2pixHD framework.

\subsection{Conditional GANs}
As described in \cite{gan}, GANs are generative adversarial networks, which consist of two adversarial models: the generator and the discriminator ($G$ and $D$). The generator network learns to map noise variables $z \sim  P_{Noise(z)} $ to  a  complex  distribution $x=G(z)$, where $z$ is the noise sample and $x$ is the data sample.  The generative model $G$ attempts to make the generated sample distribution $P_{G}(x)$  indistinguishable from the actual data distribution $P_{Data}(x)$. The discriminative model $D$ is trained to identify the $real$ samples from the data sample and the $fake$ samples from $G$. Here $D$ and $G$ play a min-max game, which can be represented in a value function:
\begin{multline}
\mathcal{L}_{GAN}(G, D)=\min\limits_{G}\max\limits_{D}\mathbb{E}_{x\sim P_{Data}(x)}[logD(x)]\\ + \mathbb{•}{E}_{z\sim P_{Noise}(z)}[log(1 - D(G(z))) \label{gan}
\end{multline}
Due to the weak guidance of the generative model, the generated samples cannot be controlled. Therefore, the conditional GAN is proposed to guide the generation by considering additional information $y$. The training objective of conditional GAN can be expressed as:
\begin{multline}
\mathcal{L}_{cGAN}(G, D)=arg\min\limits_{G}\max\limits_{D}\mathbb{E}_{x\sim P_{Data}(x)}[logD(x|y)]  \\ 
+ \mathbb{E}_{z\sim P_{Noise}(z)}[log(1 - D(G(z|y)|y))] \label{cgan}
\end{multline}

\subsection{Pix2pixHD}

Pix2PixHD framework, improved the Pix2Pix \cite{pix2pix} framework and achieved good results in image translation by using:
\begin{enumerate}
\item A coarse-to-f\mbox{}ine generator. \emph{Coarse}: residual global generator network ($G1$) trained on lower resolution images. \emph{F\mbox{}ine}: another residual  local enhancer network ($G2$) appended to $G1$ and the two networks are trained jointly the images.
\item Multi-scale discriminator architectures have identical network structures but operate at dif\mbox{}ferent image scales.

\item A robust adversarial learning objective function, that is based on the conditional GAN loss function shown in Eq.\ \ref{cgan}, combines with the feature matching loss to stabilize the training process, as the generator has to produce realistic image features on dif\mbox{}ferent scales. The feature matching loss $\mathcal{L}_{FM}(G, D_{k})$ is given by:
\begin{multline}
\mathbb{E}_{(s, x)\sim P_{Data}(s, x)}\sum_{i=1}^{T}\dfrac{[||D_{k}^{i}(s, x)-
D_{k}^{i}(s,G(s))||_{1}]}{N_{i}}  \label{fm}
\end{multline}
where $s$ is the conditional information semantic label map, $x$ is a corresponding original image, $T$ is the total number of layers and $N_{i}$ denotes the number of elements in each layer.
\end{enumerate}
The proposed enhancement methods make the conditional GANs able to synthesize high-resolution image samples.
\section{Super-resolution mel-spectrogram}
\begin{figure}[t]
  \centering
  \includegraphics[width=\linewidth]{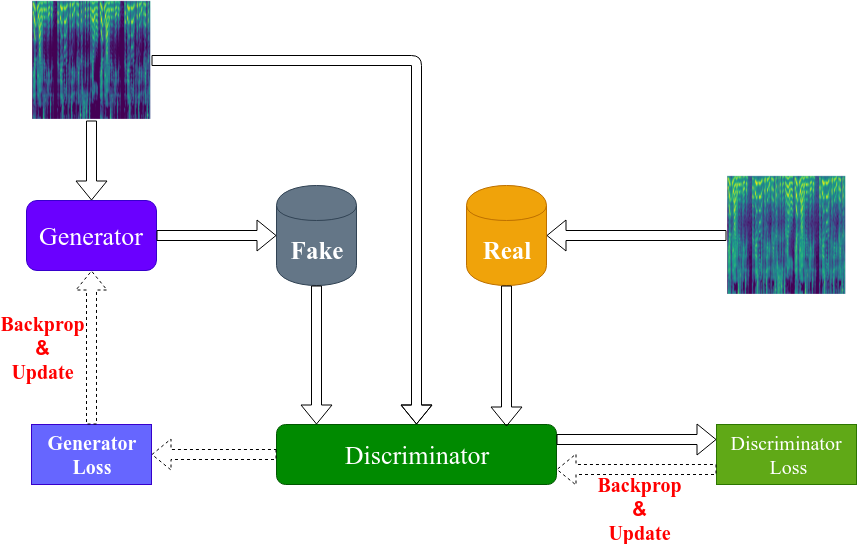}
  \caption{Illustration of cGAN system architecture.}
  \label{fig:cgan}
\end{figure}

\begin{figure}[!htbp]
\centering
\subfigure[Components of the ResUnet architecture]{
\begin{minipage}[t]{\linewidth}
\centering
\includegraphics[width=\linewidth]{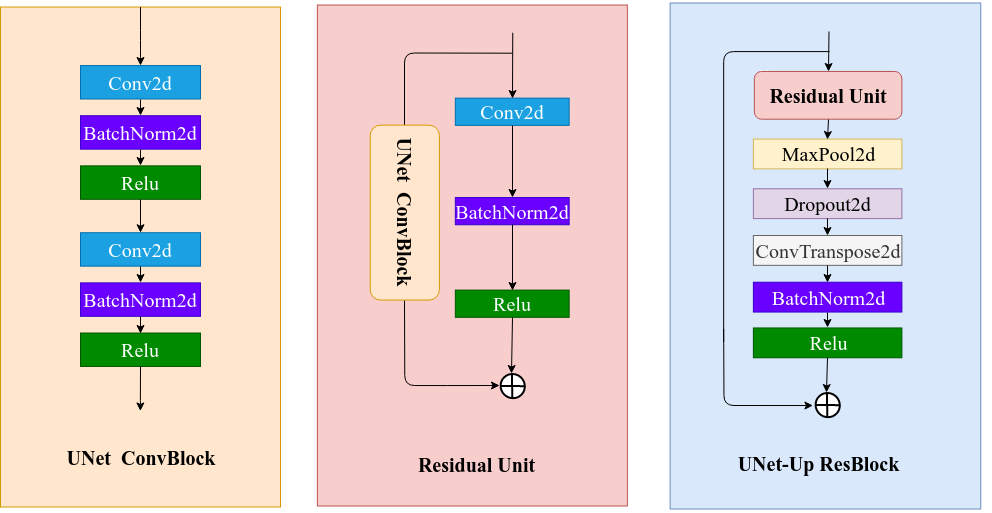}
\end{minipage}%
}%

\centering
\subfigure[ResUNet with long-skip residual connection]{
\begin{minipage}[t]{\linewidth}
\includegraphics[width=\linewidth]{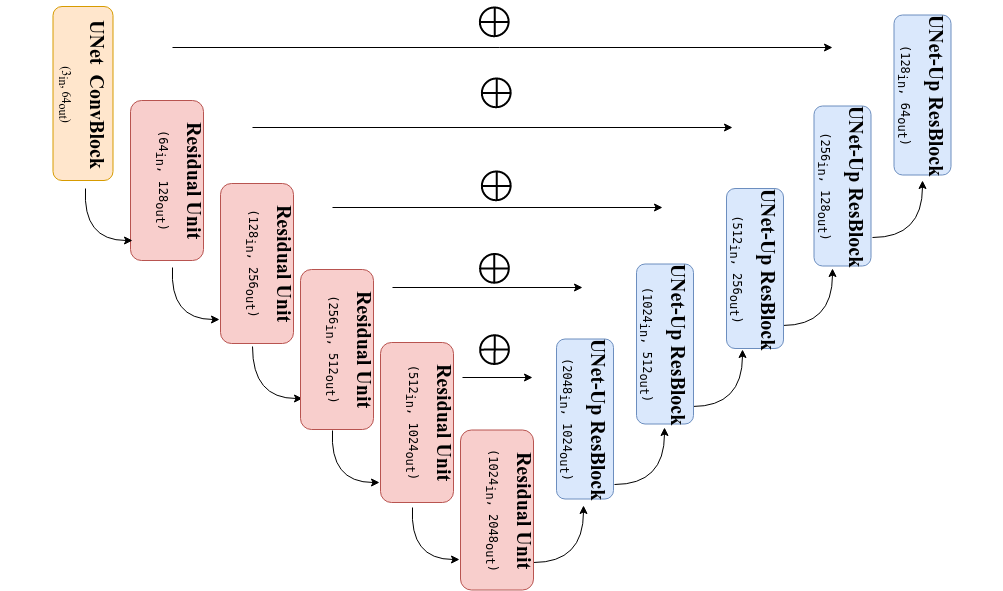}
\end{minipage}%
}%

\centering
\subfigure[Downsampling and upsampling blocks]{
\begin{minipage}[t]{\linewidth}
\includegraphics[width=\linewidth]{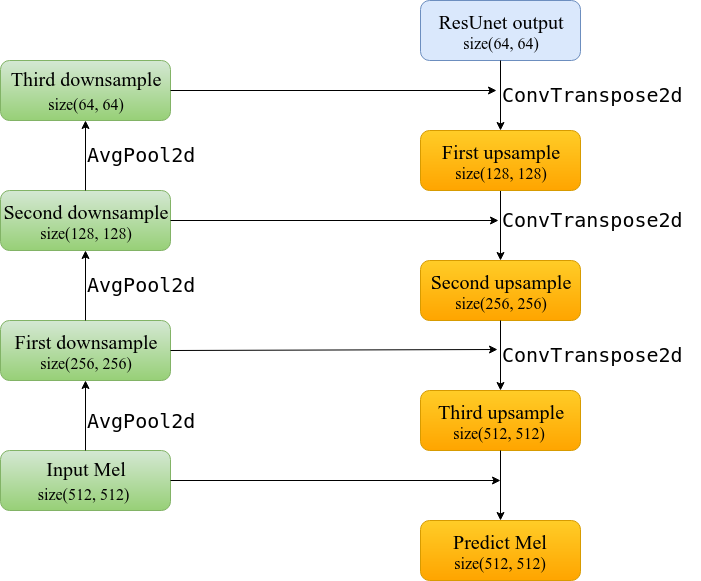}
\end{minipage}%
}%

\centering
\caption{Local enhancer generator architecture}
\label{fig:resunet}
\end{figure}

The goal of generating super-resolution mel-spectrogram is to produce a high-resolution synthetic mel-spectrogram with f\mbox{}ine-grained details for a given coarse mel-spectrogram (as shown in F\mbox{}ig.\ \ref{fig:cgan}). We propose to utilize the adversarial strategy to train a ResUnet (as shown in Fig.\ \ref{fig:resunet}) that updates the generator of Pix2PixHD. The components of ResUnet and its contrasts with Pix2PixHD are explained in the flowing paragraphs.

\begin{figure}[t]
  \centering
  \includegraphics[width=1\linewidth]{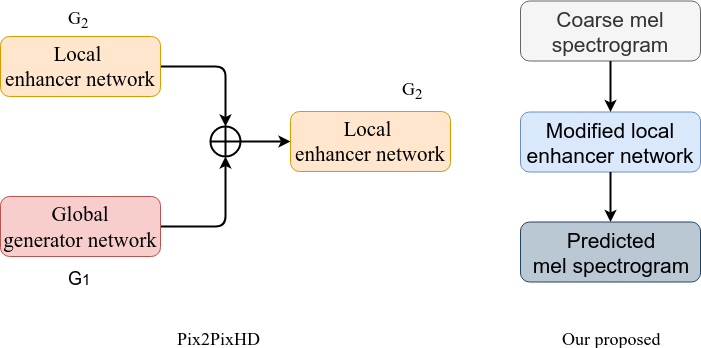}
  \caption{The differences between the Pix2PixHD and our model: (a) a modified local enhancer network. (b) The global generator network is not used in our model.}
  \label{fig:g2}
\end{figure}

\subsection{Local enhancer generator}
Difference from Pix2PixHD, which decomposes the generator into two sub-networks ($G1$ and $G2$), shown in Fig. \ref{fig:g2}. The $G2$ is only required as the input mel-spectrograms similar to the real mel-spectrograms. Then we are required to make partial enhancements to restore the lost information. ResUnet is mainly used to build and restructure information and then represent the images in a local neighborhood image space. Our local enhancer network consists of a ResUnet with a set of residual downsampling and upsampling blocks, shown in Fig. \ref{fig:resunet} 

\subsection{Adversarial loss}
In Pix2PixHD, the full objective function is to achieve multi-scale discriminator loss and feature matching loss. We use four multi-scale discriminators $D = (D_1, D_2, D_3, D_4)$ to distinguish the real mel-spectrogram from the generated ones, using the loss function in Eq.\ \ref{Multi-d} . In addition, we also add Structural Similarity Index (SSIM)\ \cite{ssim} loss and Mean Squared Error (MSE) loss to increase the stability of the training.
\begin{equation}
\mathcal{L}_{Adversarial}=\min_G\max_{D_1,D_2,D_3,D_4}\sum_{k=1}^4\mathcal{L}_{GAN}(G,D_k) \label{Multi-d}
\end{equation}

SSIM is one of the state-of-the-art perceptual metrics to measure image quality, which has been shown to provide sensitivity to structural information and texture.  SSIM is def\mbox{}ined between $0$ and $1$, where $1$ indicates perfect perceptual quality relative to the ground truth.  For pixel $p$, the SSIM value is computed as:
\begin{equation}
SSIM(p) = \frac{2\mu_x\mu_y + C_1}{\mu_{x}^2+\mu_y^2+C_1} \cdot \frac{2\sigma_{xy} + C_2}{\sigma_x^2+\sigma_y^2+C_2} \label{ssim-value}
\end{equation}
where $\mu_x$ and $\mu_y$ are the mean for regions $x$ and $y$; $\sigma_x$ and $\sigma_y$ are standard deviation for regions $x$ and $y$; $\sigma_{xy}$ is the covariance of regions $x$ and $y$ and $C_1$ and $C_2$ are constant values for stabilizing the denominator. The SSIM loss for every pixel $p$ is expressed as:
\begin{equation}
\mathcal{L}_{SSIM}= 1- SSIM(p) \label{ssim}
\end{equation}

The MSE loss, used for image restoration tasks is def\mbox{}ined as:
\begin{equation}
\mathcal{L}_{MSE}= \frac{1}{N_p}\sum_{p=0}^{N_p}(x(p)-y(p))^2 \label{mse}
\end{equation}
where $p$ denotes the pixel and $N_p$ denotes the total number of pixels in images $x$ and $y$.

Our full objective function is defined as the combinations of two parts: (1) the GAN and feature matching loss. (2) the intensity loss MSE and the structural loss SSIM. This is expressed as:
\begin{multline}
\mathcal{L}_{Final} = \alpha(\mathcal{L}_{Adversarial} + \mathcal{L}_{FM}) + (1-\alpha)(\mathcal{L}_{SSIM} + \mathcal{L}_{MSE})
\end{multline}
where $\alpha$ is a constant that controls the importance factor of the loss terms.

\section{Experiments}
\subsection{Corpus and features}
Our work uses an open source LJSpeech dataset \cite{ljspeech}, which  consists of $13,100$ pairs of text and short audio  clips of a single English speaker. The speech waveforms have a sampling frequency of \,$22050$ Hz, duration of $24$ hours and without phoneme-level alignment. We randomly selected $13,000$ utterances for training networks and another $100$ for testing data.

The mel-spectrograms are extracted from the original speech data with window length 1024, hop length 256 and frame length of 1024 as the parameters of fast Fourier transform (FFT) . To obtain the mel-spectrogram inputs (baseline), we use Grif\mbox{}f\mbox{}in-Lim algorithm to invert the original mel-spectrograms into time-domain waveforms, from which we use the same parameters to extract the input mel-spectrograms. We note that Grif\mbox{}f\mbox{}in-Lim algorithm produces characteristic artifacts and low quality audio, which is a common problem of synthesizing speech as it creates a gap between the input and output mel-spectrograms.

Another idea is to take the mel-spectrogram as an image. The stored mel-spectrogram adopts a series of scaling transformations that are reversible.  The size of our extracted mel-spectrogram is $80\times785$\ ($80$ is the Mel f\mbox{}ilterbank channel, $785$ is the number of frames in the mel-spectrogram). Due to limitation of computing resources, we scaled the mel-spectrograms of model input to size $512\times512$. If we upscale to $1024\times1024$, we may obtain better performance than an other scale setting. 
%
%

\subsection{Vocoder methods}
In our experiments, we use Grif\mbox{}f\mbox{}in-Lim algorithm and WaveNet vocoder to synthesize speech from mel-spectrograms and then evaluate the quality of the synthesized speech. If the mel-spectrogram generated by our proposed model can restore the information lost by the coarse mel-spectrogram, better quality of synthesized speech maybe achieved.
\subsubsection{Grif\mbox{}f\mbox{}in-Lim}
Grif\mbox{}f\mbox{}in-Lim algorithm is based on iteratively estimating the missing phase information and converting between modif\mbox{}ied magnitude spectrograms and actual signal spectrograms. The iterative process may degrade the quality of synthesized speech. For simplicity of Grif\mbox{}f\mbox{}in-Lim algorithm, we use $60$ iterations from frequency to time domain in our experiments.
\subsubsection{WaveNet}
WaveNet is a typical autoregressive generative model with a convolutional model architecture with dilated convolutions that learns directly to model raw signals in the time domain and achieves very high-quality synthetic speech. The log-scale mel-spectrogram has been found to be a good acoustic feature as it reconstructs a more accurate algorithm than Grif\mbox{}f\mbox{}in-Lim. 

\subsection{Model setup}
We trained our model using $4$ NVidia GeForce GTX TITAN X GPUs with $13000$  $<$coarse mel-spectrogram, original mel-spectrogram$>$ image pairs as input, batch size of $16$ and $100$ epochs for $4$ days. We used Adam optimizer \cite{adam} with $\beta_1=0.5$, $\beta_2=0.999$, $\epsilon=10^{-6}$ and an initial learning rate of $2\times10^{-4}$.\ After $40$ epochs, we started a linearly decaying learning rate which decays to $0$. We f\mbox{}irst trained the local multi-scale enhancer network with two-scale, three-scale and four-scale, and found that when we adopt four-scale, we can ef\mbox{}fectively accelerate loss function decline. Correspondingly, $4$ discriminators were used with identical network structure at dif\mbox{}ferent image scales.
\subsection{Evaluation}
To  evaluate  the  quality  of  the  enhanced  speech, we used Grif\mbox{}f\mbox{}in-Lim algorithm and pre-trained WaveNet model. The pre-trained model available at \\\texttt{https://github.com/r9y9/wavenet\_vocoder} to synthesize speech from predicted, coarse and original mel-spectrograms. 
\subsubsection{Objective Evaluation}
\begin{figure}
\centering
\includegraphics[height=8cm, width=9cm]{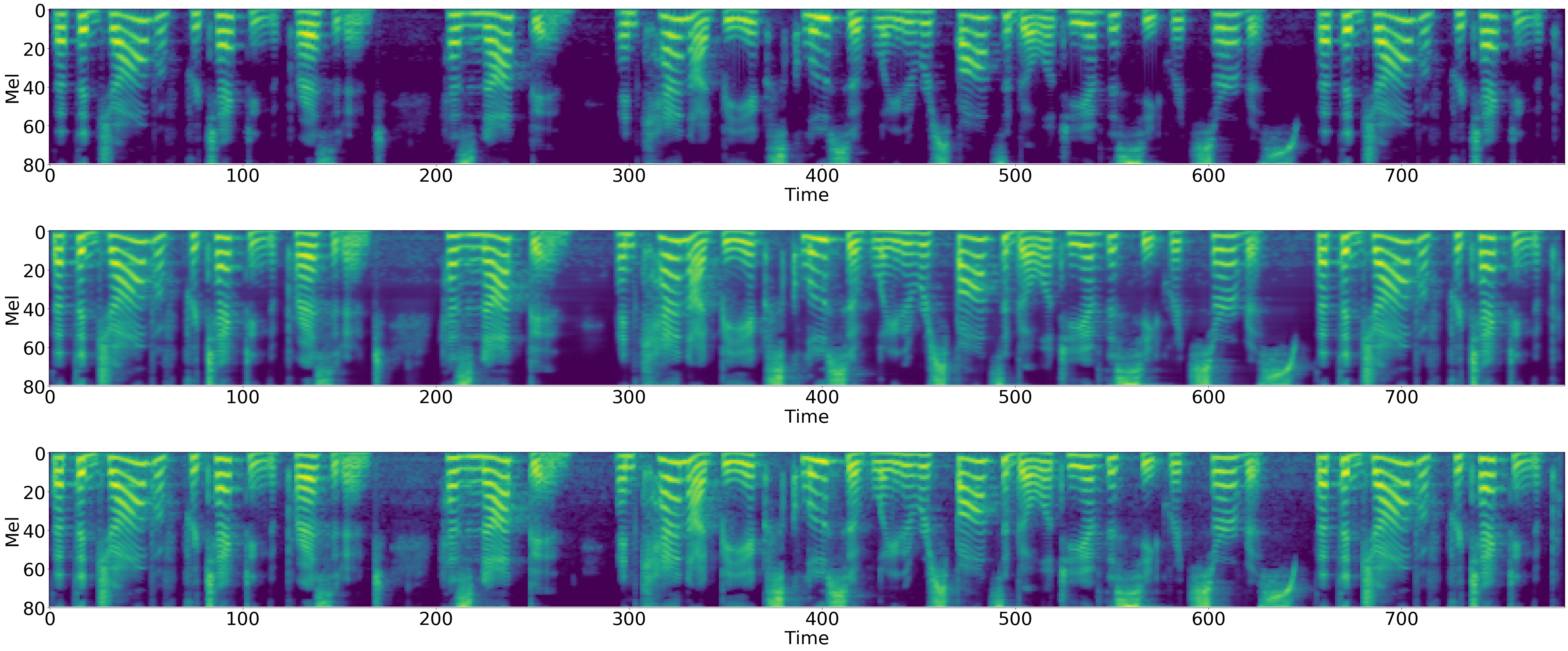}
\centering
\caption{ Comparison of Coarse mel-spectrogram (top), Predicted mel-spectrogram (middle) and Original mel-spectrogram (bottom).}
\label{fig:result}
\end{figure}

We selected mel-spectrograms generated by our model from the test dataset and compared it with the coarse mel-spectrograms and original mel-spectrograms as shown in 
Fig.\ \ref{fig:result} . We observed that our model can  not  only  emphasize  the  harmonic  structure,  but can reproduce the detailed structures that are close to those in the original mel-spectrograms from the coarse spectrograms. Moreover, we utilize SSIM value as a combined image quality measure of mel-spectrogram and Short-Time Objective Intelligibility (STOI) \cite{stoi} index as a measure of speech intelligibility. Table  \ref{ssim} shows the enhanced mel-spectrogram has a significant improvement, and table \ref{stoi} shows the generated speech from predicted mel-spectrogram and original is highly correlated with the intelligibility. We also found that the SSIM and STOI values are relatively well correlated.

\begin{table}[t]
  \caption{Mel-spectrogram comparison in terms of SSIM}
  \label{tab:ssim}
  \centering
  \begin{tabular}{ll}
    \toprule
    \textbf{Mel-spectrogram } & \textbf{ SSIM}   \\
    \midrule
    Coarse (baseline)                    & $0.591$     \\
    Predicted (ours)         & \textbf{$0.920$}  \\
  Original              & $1.00$      \\
    \bottomrule
  \end{tabular}
   \label{ssim}
\end{table}

\begin{table}[t]
  \caption{Speech intelligibility comparison in terms of STOI }
  \label{tab:stoi}
  \centering
  \begin{tabular}{lll}
    \toprule
    \textbf{ Mel-spectrogram} & \textbf{Grif\mbox{}f\mbox{}in-Lim}  & \textbf{WaveNet}   \\
    \midrule
    Coarse (baseline)                    & $0.791$      & $0.788$     \\
    Predicted (ours)         & \textbf{$0.978$}       & $0.919$    \\
  Original              & $1.00$    & $1.00$     \\
    \bottomrule
  \end{tabular}
   \label{stoi}
\end{table}

\subsubsection{Subjective evaluation}
The commonly used Mean Opinion Score (MOS) test was conducted to compare the  synthesized audio sample from coarse mel-spectrograms with the predicted mel-spectrograms. To make sure the results are legitimate, we used $100$ unseen coarse mel-spectrograms to generate the predicted mel spectrograms. We then used Grif\mbox{}f\mbox{}in-Lim and WaveNet vocoder to synthesize speech and wherein sent to Amazon's Mechanical Turk human rating service, having raters listen to the audio and rate on a f\mbox{}ive-point scale $0.5$ point increment. Our samples were rated by $8$ native listeners and the subjective MOS was calculated.
 
The results of subjective MOS are shown in Table \ref{mos} with $95\%$ conf\mbox{}idence intervals computed from t-distribution for various systems.  As we can see, our best model produced better mel-spectrograms than the coarse mel-spectrograms (baseline). The absolute improvement, obtained by using Grif\mbox{}f\mbox{}in-Lim as vocoder was $0.44$ and the one obtained by using WaveNet as vocoder was $0.17$, both improving the quality of synthesized speech. The mel-spectrograms predicted by our model achieves a subjective MOS, comparable the original mel-spectrogram regardless of whether it is synthesized by Grif\mbox{}f\mbox{}in-Lim or WaveNet. Audio samples available at  \url{https://speech-enhancer.github.io/}.
\begin{table}[t]
  \caption{Mean Opinion Scores}
  \label{tab:word_styles}
  \centering
  \begin{tabular}{lll}
    \toprule
    \textbf{Mel-spectrogram} & \textbf{Grif\mbox{}f\mbox{}in-Lim}  & \textbf{WaveNet}   \\
    \midrule
    Coarse (baseline)                    & $3.29 \pm 0.069$      & $3.84 \pm 0.061$     \\
    Predicted (ours)         & \textbf{$3.73 \pm 0.055$}       & $4.01 \pm 0.055$    \\
    Original              & $3.74 \pm 0.054$    & $4.04 \pm 0.056$     \\
 
    \bottomrule
  \end{tabular}
  \begin{tablenotes}
      \small
      \item Note: ground truth:	$4.23\pm 0.046$
    \end{tablenotes}
   \label{mos}
\end{table}

\section{Conclusions}
In our work, we have proposed the image-to-image translation model for high-quality speech synthesis from mel-spectrograms. The proposed model combines the advantages of Pix2PixHD and ResUnet (residual learning and U-Net). The multi-scale local enhancer network and skip connections within the residual units can reproduce detailed structures of mel-spectrograms. Our method signif\mbox{}icantly outperforms the baseline under subjective listening tests in MOS. But the results also showed a certain gap between the synthesized speech and the ground truth, shown by subjective listening tests in MOS. This may be caused by limitations of the vocoder. In the future work, we will design better vocoder to tackle the problem.


\bibliographystyle{IEEEbib}

\bibliography{mybib}

\end{document}